\documentclass[aps,prb,twocolumn,groupedaddress]{revtex4-1}
\usepackage[T1]{fontenc}
\usepackage[utf8]{inputenc}
\usepackage{amssymb}
\usepackage{amsmath}
\usepackage[switch]{lineno}
\usepackage{graphicx}
\usepackage{footmisc,booktabs}

\begin{document}
\title{Phonon-drag effect in FeGa$_3$}

\author{Maik Wagner-Reetz}
\author{Deepa Kasinathan}
\author{Walter Schnelle}
\author{Raul Cardoso-Gil}
\author{Helge Rosner}
\author{Yuri Grin}
\affiliation{Max-Planck-Institut f\"{u}r Chemische Physik fester Stoffe\\N\"{o}thnitzer Stra{\ss}e 40, 01187 Dresden, Germany}

\author{Peter Gille}
\affiliation{Ludwigs-Maximilians-Universit\"{a}t M\"{u}nchen\\Theresienstra{\ss}e 41/II, 80333 M\"{u}nchen, Germany}

\begin{abstract}
The thermoelectric properties of single- and polycrystalline FeGa$_3$ are systematically investigated over a wide
 temperature range. At low temperatures, below 20\,K, previously not known pronounced peaks in the thermal conductivity (400-800\,W\,K$^{-1}$\,m$^{-1}$) with corresponding 
 maxima in the thermopower (in the order of -16000\,$\mu$V\,K$^{-1}$) were found in single crystalline samples.  Measurements in single crystals along [100] and [001] 
 directions indicate only a slight anisotropy in both the electrical and thermal transport.  
From susceptibility and heat capacity measurements, a magnetic or structural phase transition was excluded.
Using density functional theory-based calculations, we have 
revisited the electronic structure of FeGa$_{3}$ and compared the magnetic (including correlations) and non-magnetic electronic
 densities of states. Thermopower at fixed carrier concentrations are calculated using semi-classical Boltzmann transport
 theory, and the calculated results match fairly with our experimental data. 
The inclusion of strong electron correlations treated in a mean-field manner (by LSDA+$U$) does not improve 
this comparison, rendering strong correlations as the sole explanation for the low temperature enhancement unlikely.
Eventually, after a 
 careful review, we assign the peaks in the thermopower as a manifestation of the phonon-drag effect, 
 which is supported by thermopower measurements in a magnetic field. 
\end{abstract}

\maketitle

\section{introduction}

Fe-based semiconductors like FeSi, FeSb$_{2}$ and FeGa$_{3}$ have attracted much attention in the past years due to their interesting physical properties. The hybridization of the transition metal $d$ orbitals
with the main group metal $p$ orbitals opens up small band gaps of the order of 0.1\,eV for FeSi,\cite{sales1994} 0.03\,eV for FeSb$_{2}$\cite{petrovic2003} and 0.47\,eV for FeGa$_{3}$.\cite{hadano_thermoelectric_2009} Unusual transport properties and their origins are strongly debated within the community for the past two decades. For example, no clear consensus has been reached on the origin of the extremely high Seebeck coefficient of -45000\,$\mu$V\,K$^{-1}$ around 10\,K in FeSb$_{2}$ single crystals.\cite{bentien2007} Until now the community is split between two main schools of thought, one favoring	the presence of strong electron-electron correlations\cite{bentien2007,sun_narrow_2009,sun2011} while the other proposes phonon-drag mechanism\cite{takahashi2011,pokharel2013} as the source for the colossal value of Seebeck effect. Herein, we present for the first time huge thermopower values in the order of -16000\,$\mu$V\,K$^{-1}$ below 20\,K for single crystals of FeGa$_{3}$. The experimental results are based on especially good quality single crystals whose chemical compositions have been carefully checked by X-ray powder diffraction and metallographic analysis. Among the various reasons that can rationalize such a large Seebeck coefficient, the main candidates are
(i) the presence of strong electronic correlations, (ii) structural phase transitions, (iii) magnetic phase transitions, and (iv) the interaction of the phonons with the mobile charge carriers via the phonon-drag effect.
For example, 
most of the low-temperature anomalies including the enhanced low temperature thermopower in FeSi are explained upon including the electronic correlations in
realistic many-body calculations.\cite{tomczak_signatures_2012} 
The onset of a magnetic phase transition has been demonstrated as the reason for the presence of distinct 
features in thermopower in Mn$_{2-x}$Cr$_{x}$Sb.\cite{wijngaard1992} 
Unusual jumps in Seebeck coefficient data have been measured in 
the parent compound of the Fe-based superconductor LaFeAsO,\cite{mcguire2008} Zn$_{4}$Sb$_{3}$\cite{souma2002} and La$_{1.85-x}$Nd$_{x}$Sr$_{0.15}$CuO$_{4}$,\cite{buechner1993} all of which have been
explained in terms of structural phase transitions.
Recently, the observation of a large negative Seebeck coefficient of -4500\,$\mu$V\,K$^{-1}$ at 18\,K in single crystalline CrSb$_{2}$ has been suggested to emerge from the dominating 
phonon-drag effect at low temperatures.\cite{sales2012} 

The intermetallic compound FeGa$_{3}$ was first found during the systematic studies in the binary 
system Fe-Ga. The crystal structure was originally ascribed to a non-centrosymmetric structure type 
IrIn$_{3}$.\cite{schubert1959,schubert1958,dasarathy1965} On the contrary, a centrosymmetric space group
$P$4$_{\mathrm{2}}/mnm$ was suggested later for the description of the symmetry for the crystal structure
of FeGa$_{3}$.\cite{lu1965} More recent studies confirmed the centrosymmetric crystal structure and 
reported the first observation of semiconducting behavior.\cite{haussermann_fega3_2002} 
Thermoelectric properties (TE) at temperatures from 300 to 950\,K with negative thermopower values were later determined on polycrystalline samples.\cite{amagai_thermoelectric_2004} 
 The first characterization on FeGa$_{3}$ single crystals depending on the crystallographic orientation was reported in Ref.~\onlinecite{hadano_thermoelectric_2009} but the authors observe the appearance of gallium inclusions of approximately 3\,\% in their samples. FeGa$_{3}$ was found to be diamagnetic, which was later validated.\cite{tsujii_observation_2008}
 Presence of an energy gap was established using magnetic susceptibility measurements (0.29-0.45\,eV) and
 as well as by valence band X-ray photoemission spectroscopy ($\leq$ 0.8\,eV), while
 $^{57}$Fe M\"{o}ssbauer spectroscopy revealed the absence of magnetic ordering.\cite{tsujii_observation_2008}
Nevertheless, considering the narrow 3$d$ bands, Yin and Pickett\cite{yin_evidence_2010} explored the option of a magnetically ordered ground state
in FeGa$_{3}$ using total energy calculations and obtained an antiferromagnetically ordered spin-singlet state upon inclusion of correlation effects. Though no verifiable experimental evidence for a magnetic transition in FeGa$_{3}$ exists, muon spin rotation spectra\cite{storchak_thermal_2012} show the existence of a spin-polaron band, which is consistent with
the presence of an antiferromagnetic spin-singlet scenario. 
Photoemission and inverse photoemission spectroscopic experiments\cite{arita_electronic_2011} have also been performed on single crystals of FeGa$_{3}$ to
probe its electronic structure and provide an estimate for the amount of correlation effects in this material. 
The measured band dispersions were comparable to calculations based on density functional theory (DFT) which included a Coulomb repulsion parameter $U$ $\approx$ 3\,eV. Nonetheless, the authors conclude that the correlation
effects in FeGa$_{3}$ is not as significant as in FeSi. 
Recently, another study on doped Fe(Ga,Ge)$_{3}$ systems using total energy calculations reports
an itinerant mechanism ($i.e.$ suggesting a minor role of the electron correlations implicitly) for the experimentally observed ferromagnetism in the $n$-type samples.\cite{umeo2012,singh2013}

In our opinion, the presence/absence of magnetism and/or correlations for the stoichiometric FeGa$_{3}$ systems
is still an open question. 
Until now, there have been no reports of anomalous transport behavior at low temperatures in FeGa$_{3}$. 
We have for the first time grown well-defined single crystals of FeGa$_{3}$ without any Ga inclusions
and with a total mass up to 25\,g
and performed various thermodynamic and transport measurements. Unusually large 
Seebeck coefficient (in the order of -16000\,$\mu$V\,K$^{-1}$) and thermal conductivity (400-800\,W\,K$^{-1}$\,m$^{-1}$) below 20\,K in single crystalline samples are observed, which disappears in the polycrystalline specimen with
equivalent experimental composition. As mentioned in the opening paragraph, various scenarios could be behind these large Seebeck coefficient and thermal conductivity values. In our work, all the options are explored in detail to rationalize the observed low temperature feature.

\begin{table*}
\caption{Lattice parameters and experimental compositions from WDXS analysis for single- and polycrystalline 
samples of FeGa$_3$.\label{tab:chemie}}
\begin{ruledtabular}
\begin{tabular}{ccccl}
Experimental composition	&	\textit{a} (\AA)	&	\textit{c} (\AA)	&	\textit{V} (\AA$^3$)	&	Remarks\\
\midrule
Fe$_{1.023(4)}$Ga$_{2.98(1)}$	&	6.2661(1)	&	6.5597(3)	&	257.56(1)	&	Single crystal\\
                 						&	&	&	&	(flux-grown)\\
Fe$_{1.04(1)}$Ga$_{2.96(1)}$	&	6.2663(1)	&	6.5594(3)	&	257.56(1)	&	Polycrystalline SPS specimen\\
							&	&	&	&	(flux-grown)\\
Fe$_{1.018(3)}$Ga$_{2.982(4)}$	&	6.2661(2)	&	6.5596(4)	&	257.57(1)	&	Single crystal (Czochralski,\\
							&	&	&	&	oriented along [001])\\
Fe$_{1.013(3)}$Ga$_{2.987(4)}$	&	6.2665(1)	&	6.5586(2)	&	257.55(1)	&	Single crystal (Czochralski,\\
							&	&	&	&	oriented along [100])\\
FeGa$_{3}$                      &       6.2628(3)       &       6.5546(5)       &       257.09(4)       & Ref.~\onlinecite{haussermann_fega3_2002} \\
                                                        &       &       &       &       Powder (flux-grown) \\
FeGa$_{3}$                      &       6.262           &       6.556           &       257.07          & Ref.~\onlinecite{hadano_thermoelectric_2009}\\
                                                        &       &       &       &       Single crystal (flux-grown)\\  
\end{tabular}
\end{ruledtabular}
\end{table*}

\begin{figure*}
\includegraphics[width=\textwidth]{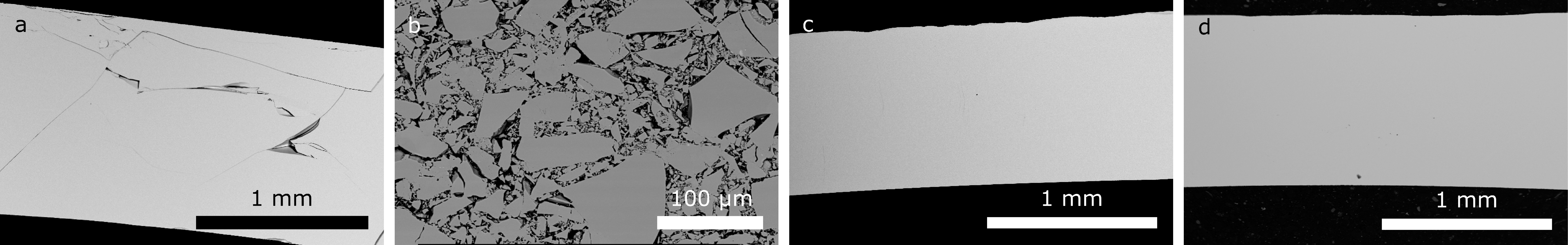}
\caption{Secondary electron microstructure pictures of FeGa$_3$ specimens: (a) flux-grown single crystal  (b) 
polycrystalline specimen after SPS of flux-grown single crystal, (c) single crystal obtained from Czochralski method and oriented along [001] and (d) single crystal obtained from Czochralski method and oriented 
along [100].\label{fig:mikro}}
\end{figure*}

\section{Methods}
\subsection{Synthesis and chemical characterization}

Single crystals of FeGa$_3$ were grown using the two-phase region FeGa$_3$ + melt of the Fe-Ga phase diagram (flux-growth). 
The elements Fe (powder, 99.99\,\%, ChemPur) and Ga (shots, 99.9999\,\%, ChemPur) in the ratio 1:20 were weighed in alumina crucibles and encapsulated 
in quartz ampoules under argon atmosphere with a pressure of 200\,mbar. The mixture was heated at a rate of 5\,K\,min$^{-1}$ 
up to 1173 K and kept at this temperature for 90 hours. After slow cooling from 1173\,K 
to 473\,K the crystals were separated from the remaining solidified melt and cleaned with hydrochloric acid to remove any residual gallium from the surface. 
 A rectangular piece, 5.8\,mm in length (longest possible rectangular bar) was cut for TE characterization.
 
The Czochralski method was used as a second technique to grow large single crystals ($\approx$ 8\,mm). Starting from a homogenized melt of composition Fe$_{16.5}$Ga$_{83.5}$ that had been prepared from Fe (bulk, 99.995\,\%, Alfa Aesar) and Ga (pellets, 99.9999\,\%, ChemPur), in a first experiment spontaneous nucleation of FeGa$_3$ occurred at a tapered corundum tip. The general approach has been described in Ref.~\onlinecite{Gille_Single_2008}. In the next crystal growth runs, use was made of [001]-oriented FeGa$_3$ seeds to grow well-defined single crystals of a total mass up to 25\,g by slowly crystallizing from a Ga-rich melt using pulling rates as low as 0.1\,mm\,h$^{-1}$ while gradually cooling down the melt according to the changing temperature of the liquid.

For polycrystalline specimens, pieces of the flux-grown crystal were ground to powder and recompacted with the
spark plasma sintering (SPS) technique. Using graphite dies, a maximum temperature of 973 K with an uniaxial pressure
of 90 MPa was applied for 10 minutes.

X-ray diffraction data sets of all samples were collected on an Image Plate 
Guinier Camera Huber G670 (Cu K$\alpha_1$ radiation, $\lambda$ = 1.54056\,\AA, 3\,$^\circ$ $\leq$ 2\textit{$\theta
$} $\leq$ 100\,$^\circ$, LaB$_6$ as internal standard, \textit{a} = 4.1569(1)\,\AA). Metallographic investigations were 
made on plain cross sections after embedding samples in conductive resin (PolyFast, Struers, Denmark) with 
additional grinding and polishing with micron-sized diamond powders. Wavelength dispersive X-ray spectroscopy (WDXS) 
investigations of the chemical composition were performed with a Cameca SX 100 spectrometer using FeSi and GaP as standards.

\subsection{Physical measurements}
Thermal diffusivity measurements of single and polycrystalline samples were 
performed with the laser flash technique (LFA 427, Netzsch, Germany) in the temperature range from 300 to 773 K. 
The thermal conductivity $\kappa$ was derived from the relation $\kappa = \alpha DC_p$, where $\alpha$ is the 
thermal diffusivity, $D$ is the density and $C_p$ is the heat capacity, respectively. The density of the obtained pellet 
determined by the Archimedes method 
reaches 92\,\% of the theoretical value. Heat capacity data at high 
temperatures were determined by differential scanning calorimetry (DSC 8500, Netzsch). The electrical resistivity and 
the Seebeck coefficient of rectangular bars were determined simultaneously in a commercial ZEM-3 equipment 
(ULVAC-Riko, Japan) in the temperature range from 300 to 773\,K.

For TE properties at low temperatures from 4 to 350\,K the thermal conductivity, electrical resistivity and Seebeck 
coefficient were measured simultaneously using a commercial system (thermal transport option of a PPMS, Quantum 
Design, USA). The dependence of thermopower on the magnetic field (\textit{H} = 0\,T, 9\,T) was also measured. Given the fact that the thermal conductivity is very high at low temperatures, we additionally reduced the cross section (thinned) at the middle of the sample as depicted in the inset of Fig. 12 in order to maximize the heat flow, while reducing the temperature jumps at the contacts of the sample to the heater and to the heat sink.

Heat capacity measurements in zero-field (ZF) and in field with \textit{H} = 9\,T were carried out using a relaxation method (HC option, PPMS). 
Magnetic properties were measured on a SQUID magnetometer (MPMS-XL-7, Quantum Design). Zero-field cooling (ZFC, measured in 
warming) and field cooling magnetization data were taken in various fields up to \textit{H} = 7\,T.

\subsection{Calculational details}

The total energy calculations to obtain the density of states (DOS)
and band structure were performed using the full potential nonorthogonal
local orbital code (\texttt{FPLO})\cite{fplo1} employing the local density approximation (LDA).
The Perdew and Wang\cite{PW92} exchange
correlation potential was chosen for the scalar relativistic calculations.
Additionally, we have explored the possibility of the presence of strong correlations
in the Fe 3$d$ shell by including an on-site Coulomb repulsion $U$ via the LSDA+$U$
method, applying the ``atomic limit'' double
counting term (also referred to as ``fully localized limit'' (FLL)).
The projector on the
correlated orbitals was defined such that the trace of the occupation number
matrices represent the 3$d$ gross occupation.
The total energies were converged on a dense $k$ mesh with 2176 points (30$\times$30$\times$30) for the
conventional cell in the irreducible wedge of the Brillouin zone.
The  transport properties were calculated using the semiclassical Boltzmann transport
theory\cite{semi1, semi2, semi3} within the constant scattering approximation as implemented in \texttt{BoltzTraP}.\cite{boltztrap}
This  approximation is based on the assumption that the scattering
time $\tau$ determining the electrical conductivity does not vary  with energy on the
scale of $k_{\mathrm{B}}T$. Additionally, no further assumptions are made for the dependence of $\tau$ on the temperature.
This method has been successfully applied to many narrow band gap materials
including clathrates, RuIn$_{3}$ derivatives and as well as other intermetallic compounds and oxides.\cite{semi3,johnsen2006,kasinathan2007,xiang2007,ruin3}
Since the evaluation of transport integrals requires an accurate estimation of band velocities,
the energy bands are calculated on a fine mesh of 9126 $k$-points in the irreducible wedge of the
Brillouin zone.

\begin{figure}[t]
\includegraphics[width=\columnwidth]{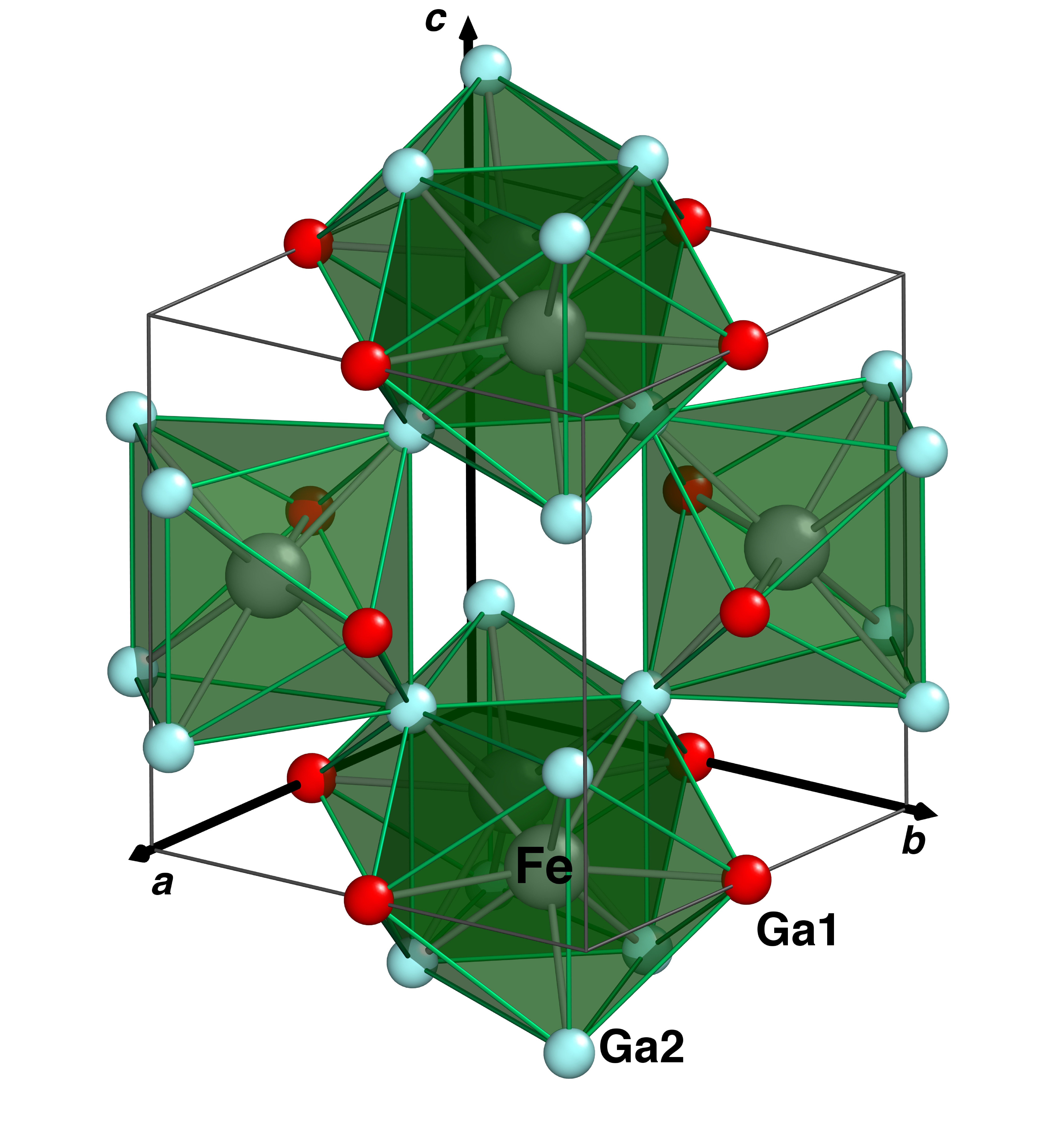}
\caption{(Color online) Tetragonal crystal structure of FeGa$_{3}$ according to Ref.~\onlinecite{haussermann_fega3_2002}. The large (grey) atoms are Fe, while the smaller (Ga1 - red, Ga2 - light-blue)
atoms are Ga [Fe at 4$f$ (0.3437,0.3437,0); Ga1 at 4$c$ (0,0.5,0); Ga2 at 8$j$ (0.1556,0.1556,0.2620)]. The polyhedra surrounding the Fe atom is composed of 8 Ga atoms: two nearest neighbor Ga at
a distance of 2.37\,\AA (Ga1),  two next nearest neighbor Ga at a distance of 2.39\,\AA (Ga2), and 
four third nearest neighbor Ga at a distance of 2.50\,\AA (Ga2).  
Each Fe atom also has one another Fe atom at a distance of 2.77\,\AA, forming so-called "dimers" in the (001) plane.
The polyhedrons of the iron atoms belonging to the same "dimer" are sharing a common face; while polyhedrons of different "dimers" are corner-sharing.
\label{fig:structure}}
\end{figure}

\begin{figure} [t]
\includegraphics[width=\columnwidth]{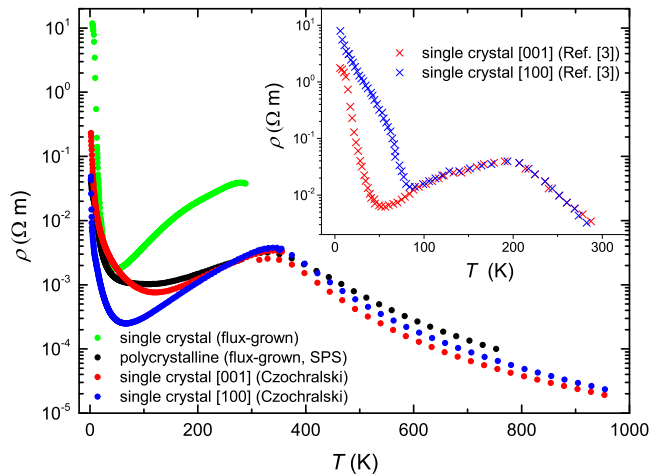}
\caption{(Color online) Electrical resistivity in dependence on the temperature for single- and polycrystalline FeGa$_3$ specimens. 
For comparison single crystal data Ref.~\onlinecite{hadano_thermoelectric_2009} are included in the inset. Band 
gap calculation with the Arrhenius law $\rho$(\textit{T}) = exp(\textit{E}$g$/2\textit{k}$_B$\textit{T}) gives energy gaps 
of 0.38-0.51\,eV.\label{fig:rho}}
\end{figure}

\begin{figure} [t]
\includegraphics[width=\columnwidth]{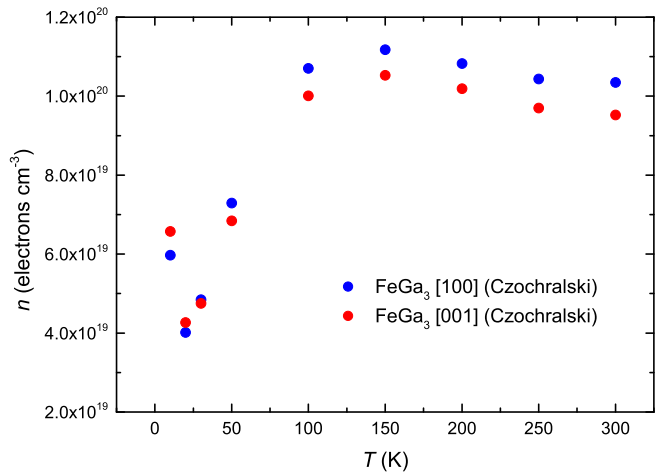}
\caption{(Color online) Charge carrier concentration $n$ as a function of temperature for the oriented FeGa$_{3}$ single crystals (Czochralski method). \label{fig:hall}}
\end{figure}

\begin{figure} [b]
\includegraphics[width=\columnwidth]{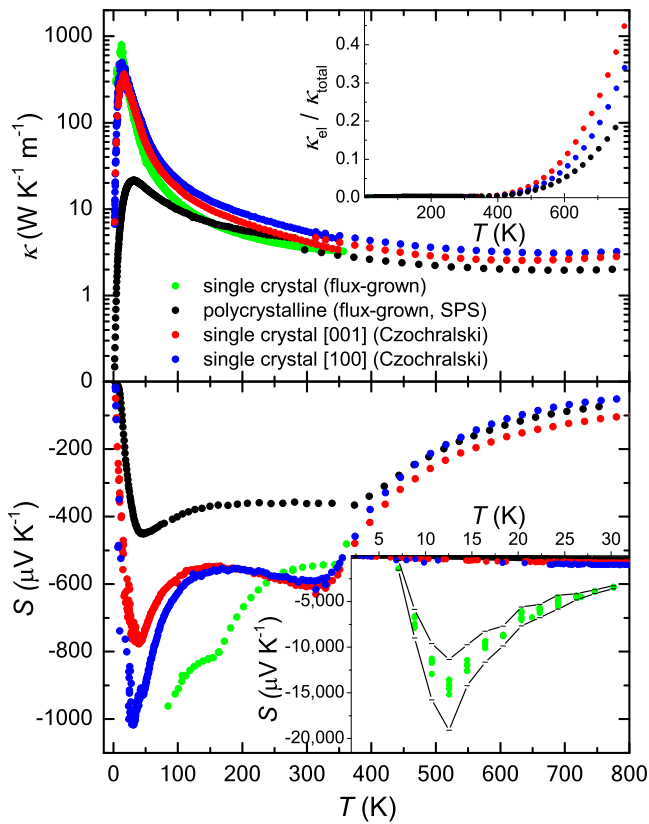}
\caption{Top: Temperature dependence of the thermal conductivity for FeGa$_3$ specimens. The inset represents the 
contribution of the electronic part to the total thermal conductivity estimated with the Wiedemann-Franz law $
\kappa_{\text{el}}$ = $\rho$\,$L_\text{0}$\,\textit{T}. Bottom: Seebeck coefficient of FeGa$_3$ specimens depending on the 
temperature. The inset shows high values for the flux-grown single crystal.
The black solid curve in the inset represents the error for the Seebeck coefficient measurement,  with $\Delta T$ = 10$^{-3}$ K.
\label{fig:kappaS}}
\end{figure}

\section{Results and discussions}
\subsection{Chemical characterizations}

According to metallographic investigations, the obtained single crystals of FeGa$_3$ via flux-growth 
and Czochralski method were free of 
non-reacted elemental iron or remaining gallium melt (Fig.~\ref{fig:mikro}).  
All X-ray diffraction patterns (See Supplemental material\footnote{See Supplemental Material at http://link.aps.org/Supplemental/... for the XRD pattern}) of the single- and polycrystalline samples 
were indexed with the tetragonal FeGa$_3$ structure type (space group $P\mathrm{4}_{\mathrm{2}}/mnm$, 
Fig.~\ref{fig:structure}) without 
any additional phases. The refined lattice parameters and chemical compositions from WDXS analysis are collected in Table~
\ref{tab:chemie}. The experimental composition for all samples are identical within the typical error limits 
of the methods. One characteristic feature is the small excess of iron atoms in comparison to the ideal composition.
In the paper, thermoelectric and thermodynamic characterizations are performed mainly
on single crystals obtained from Czochralski method.
The results can be qualitatively inferred for all single crystalline samples, since the chemical composition of
all the samples are identical. Any difference in the thermoelectric properties in polycrystalline material in 
comparison to the single crystals can be attributed  
to the changes of the microstructure. 

\subsection{Thermoelectric properties}

The logarithmic plot of the electrical resistivity as a function of temperature is shown in Fig.~\ref{fig:rho}. For all 
samples, semiconducting behavior is observed. The extrinsic region at low temperatures is characterized by high 
resistivity values implying high-purity and high-crystallinity of our materials. The flux-grown crystal 
shows higher 
resistivity with a terminated saturation range already at 280\,K, signifying the lower level of defects and 
impurities as compared to the other samples. This arises from the fact that the flux grown single
crystals were additionally cleaned with hydrochloric acid which tend to remove extrinsic impurities.
From the intrinsic region above 350\,K, band gaps were calculated with the Arrhenius law $\rho$(\textit{T}) 
= exp(\textit{E}$_g$/2\textit{k}$_B$\textit{T}), giving a value of 0.50\,eV.\footnote{The unoriented bar-shaped piece, which was cut from the flux-grown crystal could not be measured at high temperatures, owing to its short length For the unoriented crystal, a band gap of 0.38\,eV can be estimated (calculated for the intrinsic region between 280-360\,K).} The experimental data are very 
well comparable to previously published results (0.47 eV).\cite{hadano_thermoelectric_2009} A slight anisotropy in FeGa$_3$ is observed below 200\,K, with a reduced electrical resistivity in the [100] direction as compared to the [001] direction. A crossover becomes apparent around 250\,K leading to slightly 
reduced resistivity values in the [001] direction above room temperature.

To determine the carrier concentration, we measured Hall resistivity as a function of magnetic field for selected temperatures along [100]
and [001] directions for the current flow. The Hall resistivity data were fit with a linear function for both directions between 
10 and 300\,K. The determined $R_{H}$ are negative in the entire temperature range, indicating electrons being the main carriers. The carrier concentration calculated using $n$ = -1/$eR_{H}$, is plotted in Fig.~\ref{fig:hall} as a function of temperature (with $e$ = 1.602$\times$10$^{-19}$\,C). The carrier concentration does
not vary significantly between 100 and 300\,K, with room temperature values: $n^{\mathrm{[001]}}$ = 9.52$\times$10$^{19}$\,cm$^{-3}$ and $n^{\mathrm{[100]}}$ = 1.03$\times$10$^{20}$\,cm$^{-3}$. 
Below 50\,K, we observe an abrupt decrease of $n$ with the lowest value of 4$\times$10$^{19}$\,cm$^{-3}$ at 20\,K. 
Such a feature in $n$ is quite unusual and could arise from a low-mobility impurity band close to the conduction band edge, as was shown to be the case for Ge.\cite{fritzche1955} Recently, another compound where such a feature was observed is the antiferromagnetic semiconductor CrSb$_{2}$\cite{sales2012} and the authors determined an occupied low-mobility donor band at 2\,K to be causing the minimum in the carrier concentration. 

The temperature dependence of the thermal conductivity is plotted in Fig.~\ref{fig:kappaS}(a). Remarkably huge 
peaks are observed for all single crystals in the temperature range from 13--16\,K. Maximum values of 
$\approx$350\,W\,K$^{-1}$\,m$^{-1}$ for the single crystal (Czochralski method) with heat flow along [001] direction, $\approx$500\,W\,K$^{-1}$\,m$^{-1}$ for the single crystal (Czochralski method) along [100] direction and 
$\approx$800\,W\,K$^{-1}$\,m$^{-1}$ for the unoriented single crystal (flux-grown) are observed, which are in contrast to previously published data ($\approx$ 17.5\,W\,K$^{-1}$\,m$^{-1}$)  on 
oriented single crystals.\cite{hadano_thermoelectric_2009} Due to the presence of grain boundaries in 
polycrystalline FeGa$_3$, which act as additional scatterers for phonons, the peak collapses yielding 
values of only $\approx$30\,W\,K$^{-1}$\,m$^{-1}$ at 21\,K. At high temperatures, the thermal conductivity for the single crystalline 
samples differ only slightly. No significant anisotropy in thermal conduction for FeGa$_3$ is observed over the whole temperature range. 
As expected, the polycrystalline material shows the lowest thermal conductivity among all investigated samples at high 
temperatures, being consistent to already published data.\cite{amagai_thermoelectric_2004} The inset of Fig.~\ref{fig:kappaS}(a) shows the electronic contribution to the total thermal conductivity estimated via the Wiedemann-Franz law $\kappa_{\text{el}}$ = $\rho$\,$L_\text{0}$\,\textit{T}, where $L_\text{0}$ = 2.45$\times$10$^{-8}$\,W\,$\Omega$\,K$^{-2}$ is the Lorenz number. Below 400\,K the total thermal conductivity is mainly determined by the lattice thermal conductivity, having nearly no contribution from the electronic part.

The results of the thermopower measurements are depicted in Fig.~\ref{fig:kappaS}(b). In general, \textit{n}-type behavior with electrons as majority carriers is observed over the whole temperature range, which is in contrast to the isostructural compounds RuIn$_3$~\cite{Wagner_RuIn3_2011} and RuGa$_3$,\cite{WagnerReetz_RuGa3_2013} where a crossover behavior from $n$ to $p$ type is observed above room temperature. This is in line with 
the observed small Fe excess in our synthesized samples (see Table I.).\footnote{According to our DFT calculations, Fe carries a small negative charge.} At high temperatures, above 400\,K similar behavior is observed for all samples 
independent from the crystallographic orientation. Below 400\,K, distinct differences in absolute values are observed.
The polycrystalline specimen with the highest defect concentration (e.g. grain boundaries) exhibits the 
lowest minimum Seebeck coefficient, comparable to published data.\cite{amagai_thermoelectric_2004, 
haldolaarachchige_effect_2011} The slight anisotropy is presumably electronically driven with an intersection point 
of the [100] and [001] curves around 200-250\,K, as already observed in the electrical resistivity. Analogous to the 
sequence of the phonon peaks in the thermal conductivity, maximum thermopower values of -750\,$\mu$V\,K$^{-1}$ (single 
crystal via Czochralski method in [001] direction) and -1000\,$\mu$V\,K$^{-1}$ (single crystal via Czochralski method in [100] direction) appear. 
The inset of Fig.~\ref{fig:kappaS}(b) shows the maximum thermopower 
value in the order of -16000\,$\mu$V\,K$^{-1}$ measured for the unoriented single crystal (flux-grown).
Though the low temperature maximum in the thermopower occurs at the same temperature ($\approx$ 12 K)
for all single crystalline samples, the magnitude of $S$ for the flux-grown crystal is 16-20 times larger
than of the Czochralski single crystals. One main reason for this feature is the non standardized nature of
the thermal gradient arising from the length difference between the
samples: $\leq$ 6\,mm (flux-grown) vs. 8\,mm (Czochralski).
 The obtained values of thermopower are 
comparable to published data on single crystals of intermetallic CrSb$_2$\cite{sales2012} or FeSb$_2$,\cite{sun_narrow_2009} but differ significantly to orientation-dependent measurements on FeGa$_3$.\cite{hadano_thermoelectric_2009}
The authors of Ref.\,\onlinecite{hadano_thermoelectric_2009} report gallium inclusions of up to 3\% 
in their single crystalline samples, which could explain the difference between their data and our 
experimental thermopower. 

The enhancement of thermopower at low temperatures in certain narrow band gap semiconductors has been
shown to arise from increased electron-electron correlations, i.e. correlated semiconductors like Kondo insulators. 
Specifically, for iron-based semiconductors like FeSi and FeSb$_{2}$ large Seebeck coefficients at low temperatures (+500\,$\mu$V\,K$^{-1}$ at 50\,K in FeSi,\cite{sales1994} -45000\,$\mu$V\,K$^{-1}$ at 10\,K in FeSb$_{2}$\cite{bentien2007}) have been reported. This feature has been argued to arise due to the presence of strong electronic correlations in these two systems.\cite{aeppli1992,sun2011}
Though no static magnetic ordering has been observed in pure FeGa$_{3}$ until now, weak ferromagnetic
order has been observed\cite{umeo2012} for the Ge-doped FeGa$_{3-x}$Ge$_{x}$ with $x$ = 0.13 and
weakly coupled local moments have been observed for Co-doped Fe$_{1-x}$Co$_{x}$Ga$_{3}$ with $x$ = 0.05.\cite{bittar2010} These attributes in FeGa$_{3}$ are also similar to FeSi and FeSb$_{2}$ where 
no static magnetic order is observed in the pure compounds, but ferromagnetic metallic states are
discerned in FeSi$_{0.75}$Ge$_{0.25}$, FeSb$_{2-x}$Te$_{x}$ and Fe$_{1-x}$Co$_{x}$Sb$_{2}$.\cite{yeo2003,hu2009,hu2006}



\subsection{DFT calculations}

\begin{figure} [t]
\includegraphics[clip,width=\columnwidth]{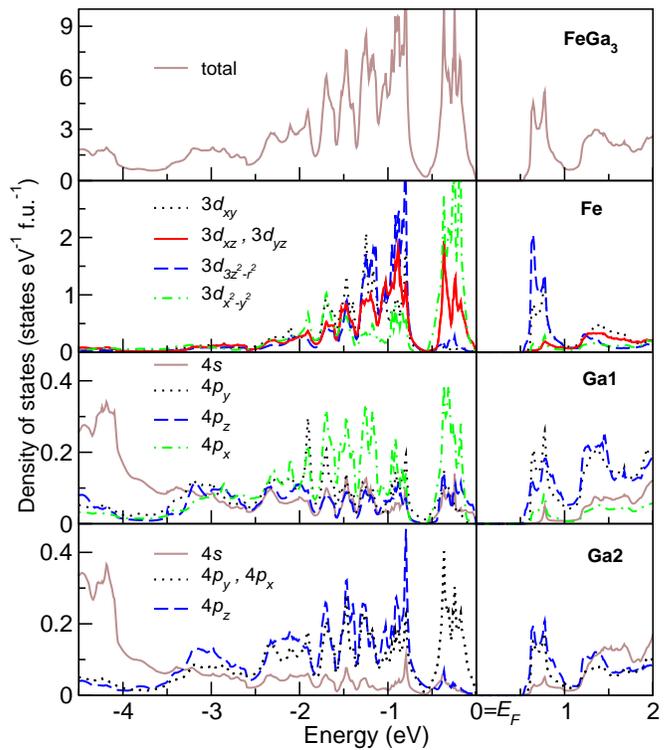}
\caption{(Color online) Non-magnetic total and site projected electronic density of states of FeGa$_3$.
\label{fig:ldabands}}
\end{figure}

\begin{figure} [b]
\includegraphics[width=\columnwidth]{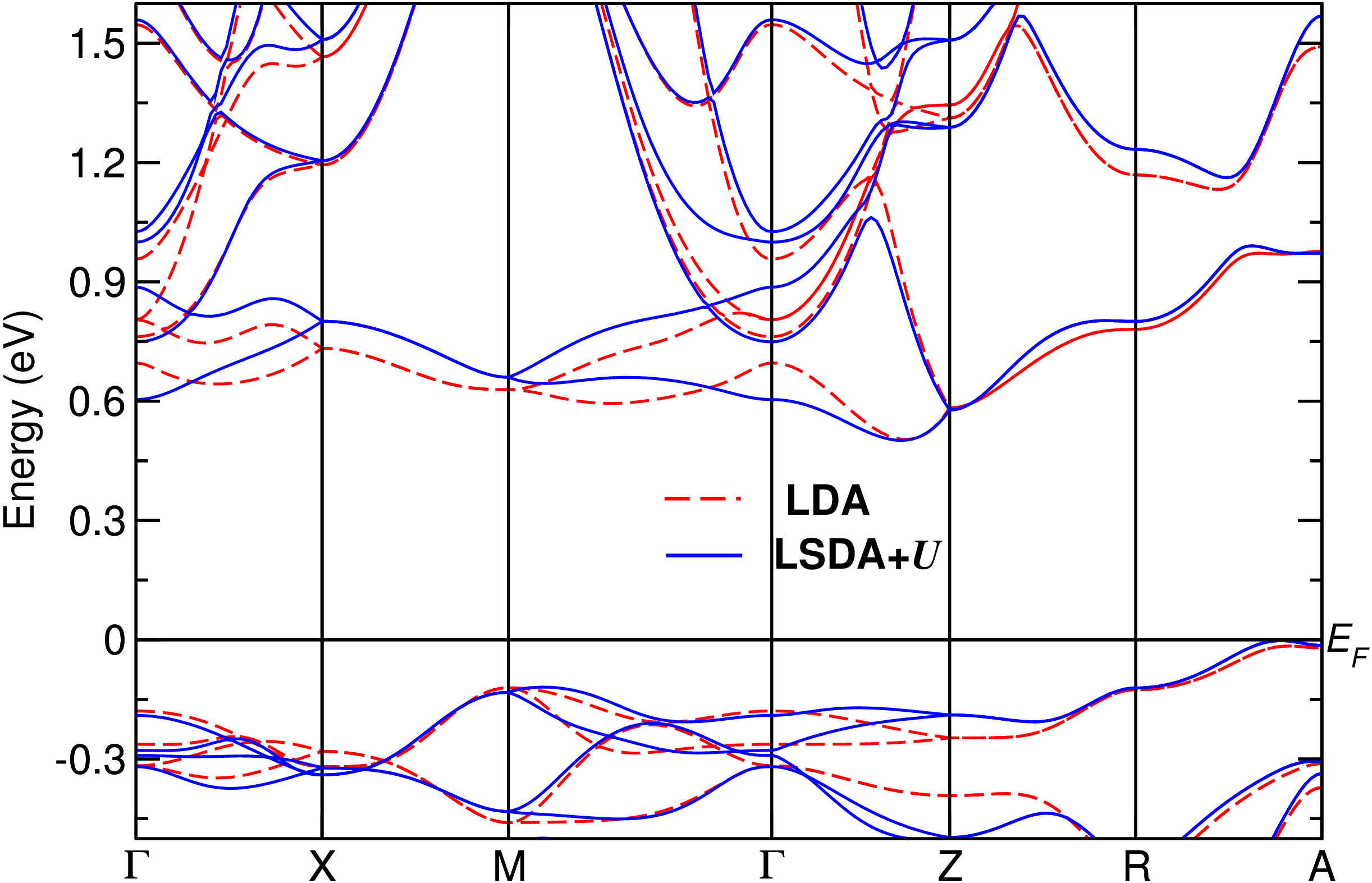}
\caption{(Color online) Comparison of the non-magnetic LDA and the AFM LSDA+$U$ band structures of FeGa$_3$.
The nearest neighbor Fe atoms are ordered antiferromagnetically in the LSDA+$U$ calculation. A $U$ value of
2 eV and a $J$ value of 0.625 eV was used. 
\label{fig:ldaubands}}
\end{figure}

\begin{figure} [t]
\includegraphics[clip,width=\columnwidth]{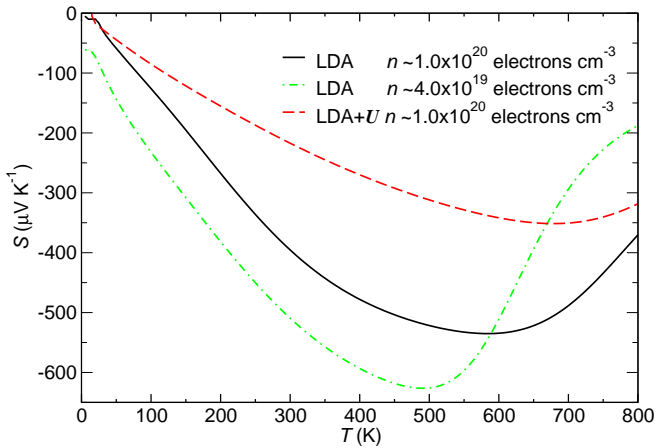}
\caption{(Color online) Calculated thermopower as a function of temperature for FeGa$_3$ for the two scenarios: LDA (non magnetic)
and LSDA+$U$ (antiferromagnetic, $U$ = 2 eV). 
There is a significant difference in the calculated thermopower between
LDA and LSDA+$U$, with the LDA curve being more in accordance with the experimentally measured thermopower.
\label{fig:thermo}}
\end{figure}

Previously, two works have addressed the eventuality of strong electronic correlations in FeGa$_{3}$ 
using DFT-based calculations, but resulted in contradictory scenarios with one paper 
suggesting the presence of strong electronic correlations in the Fe 3$d$ orbital,\cite{yin2010} while the other refutes this
scenario.\cite{guillen2012} In both these papers, the authors only consider the band structure of non-magnetic or
magnetic FeGa$_{3}$, but not the 
band structure-derived thermoelectric coefficients. Calculation of Seebeck coefficients involve the calculation
of the band velocities, which are quite sensitive to the underlying band dispersions. Since there are no 
other additional parameters involved in the calculation of the Seebeck coefficients, the results are directly
comparable to the experimental measurements, thus this could be used to identify the presence of 
strong correlations in FeGa$_{3}$. To that end, we have first calculated the 
non-magnetic LDA density of states for FeGa$_{3}$ and the results are collected in Fig.~\ref{fig:ldabands}.
Consistent with previous calculations and as well as our experiments, we obtain a semiconductor with a band gap of $\approx$
0.52 eV. The valence and conduction band edges of FeGa$_{3}$ are derived primarily from Fe $d$ states with only a 
small mixing of Ga. This feature is more clearly visible in the site projected DOS in Fig.\,\ref{fig:ldabands}. Note
the large difference in the $y$-axis values between Fe and Ga site projected DOS.  The valence band maximum 
is comprised mainly of Fe-3$d_{x^{2}-y^{2}}$ orbital character, while the conduction band minimum consists
mainly of Fe-3$d_{3z^{2}-r^{2}}$ orbital character. 
There are sharp peak-like features in the DOS on either side of the Fermi level arising from flat bands.
These sharp features are conducive for enhanced thermoelectric properties.\cite{mahan1996,freericks2003}

As mentioned previously, though no magnetism has been observed for FeGa$_{3}$ in experiments, recently there has been some discussion
about the presence of strong Coulomb correlations emerging from the narrow 3$d$ bands of this
semiconducting system using DFT based calculations. 
Using LSDA+$U$ with FLL double counting scheme, a structure wherein the nearest neighbor Fe sites are coupled antiferromagnetically with local moments on Fe of the order of 0.6$\mu_{B}$ was found to be stable.\cite{yin2010} Additionally a semiconducting gap of the order of 0.5 eV was obtained, similar to the experiments. To the contrary, using an optimized double counting scheme (screened by the Yukawa screening length $\lambda$) no magnetic moment on the iron ions was 
obtained.\cite{guillen2012} Recently, another DFT work discusses the possibility of itinerant magnetism for both Ge (on the Ga-
site) and Co (on the Fe site) doped systems.\cite{singh2013} Itinerant Stoner ferromagnetism for the doped
systems was readily obtained without invoking the need for preexisting moments in the parent semiconducting state, as well as without
the need for correlation terms. Furthermore, the Seebeck coefficients was calculated as a function of doping, though only for the non-magnetic ground state. 
In our work,  we calculate the Seebeck coefficients of FeGa$_{3}$ for both 
the un-correlated (LDA) and correlated (LSDA+$U$) scenarios to shed more light on the proceeding discussions.
Collected in Fig.~\ref{fig:ldaubands} are the results of the LSDA+$U$ calculation (with FLL double counting scheme) for $U$ = 2 eV, with a Hund's
exchange $J$ = 0.625 eV. The magnetic pattern used for this calculation is similar to that proposed in Ref.~\onlinecite{yin2010} with the two nearest Fe neighbors ordering antiferromagnetically (AFM). Surprisingly, the resulting AFM LSDA+$U$ band structure close to the Fermi edge, including the size of the semiconducting gap is quite similar to that of the non-magnetic LDA calculations.
Moving away from the Fermi edge we observe some differences between the LDA and LSDA+$U$ bands in both the
valence and conduction channels. These small differences could in turn alter the transport coefficients since they
depend on the band velocities. Significant changes in the calculated Seebeck coefficients between LDA and
LSDA+$U$ calculations could provide an alternative view for identifying the presence or absence of 
magnetism and correlations (described in a mean-field manner) in FeGa$_{3}$.

\begin{figure} [t]
\includegraphics[width=\columnwidth]{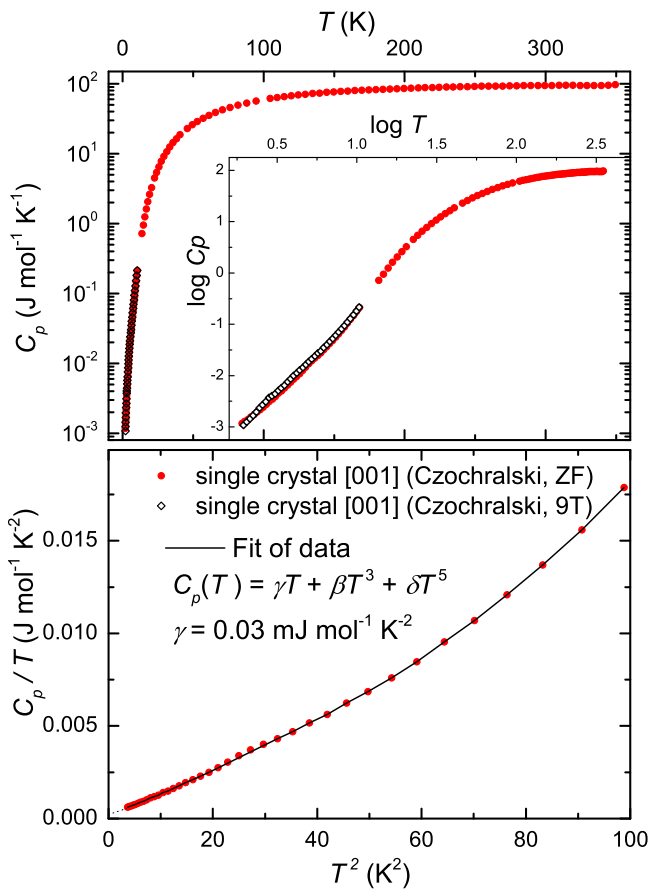}
\caption{(Color online) Top: Magnetic field dependence of the specific heat for a FeGa$_3$ single crystal (Czochralski method) oriented along [001] in 
zero-field (ZF) and high field ($H$ = 9\,T). The inset shows a log-log plot.  Bottom:
Specific heat data below 10\,K in a \textit{C}$_p$/\textit{T} vs. \textit{T}$^2$ representation resulting in a Sommerfeld 
coefficient of $\gamma$ = 0.03\,mJ\,mol$^{-1}$\,K$^{-1}$.\label{fig:Cp}}
\end{figure}

To that end, we have calculated the transport coefficients using the semiclassical Boltzmann theory. The similarity of the band gaps between LDA and AFM LSDA+$U$ calculations permits a direct comparison between the two 
scenarios. 
\footnote{We make a note about plotting the experimental
and calculated Seebeck coefficients in separate plots:  In DFT, it is instructive to calculate the Seebeck
coefficient for a fixed carrier concentration, while in reality the carrier concentration may vary as a function of temperature
during experimental measurements (see Fig.\,\ref{fig:hall}). Hence, a combined plot of the
experimental and calculated Seebeck coefficients would be misleading. 
For cases where the carrier concentration is temperature dependent,  comparison between experiment and calculation is more qualitative than quantitative. }
Collected in Fig.~\ref{fig:thermo} are calculated data of the Seebeck coefficient as a function of temperature within LDA and LSDA+$U$ for a carrier concentration $n \approx$ 1.0$\times$10$^{20}$ electrons cm$^{-3}$, obtained from Hall effect measurements for a temperature range 100-300\,K. The low temperature enhancement in the Seebeck coefficient observed in the experiments around 20\,K is not observed in either of the calculated curves. Decreasing the carrier concentration to $n \approx$ 4.0$\times$10$^{19}$ electrons cm$^{-3}$ also does not produce the low temperature feature. The Seebeck coefficient increases monotonically up to 500\,K for LDA and up to 700\,K for LSDA+$U$, after which it starts to decrease. Above 150\,K, the absolute values and the shape of $S$ within LDA are more in accordance with the experimental observations and is larger than within LSDA+$U$ by a factor of two. Though the low temperature feature is absent in our calculations, we can nevertheless infer that strong correlations, treated in a mean-field manner do not improve the description of the Seebeck coefficient in FeGa$_{3}$. 
More sophisticated calculations that treat Coulomb correlation $U$ beyond a mean-field approximation are
necessary to obtain a more in-depth understanding of the role of correlations in 
FeGa$_{3}$.\cite{Grenzebach_2006,Freericks_2001}

\begin{figure} [t]
\includegraphics[width=\columnwidth]{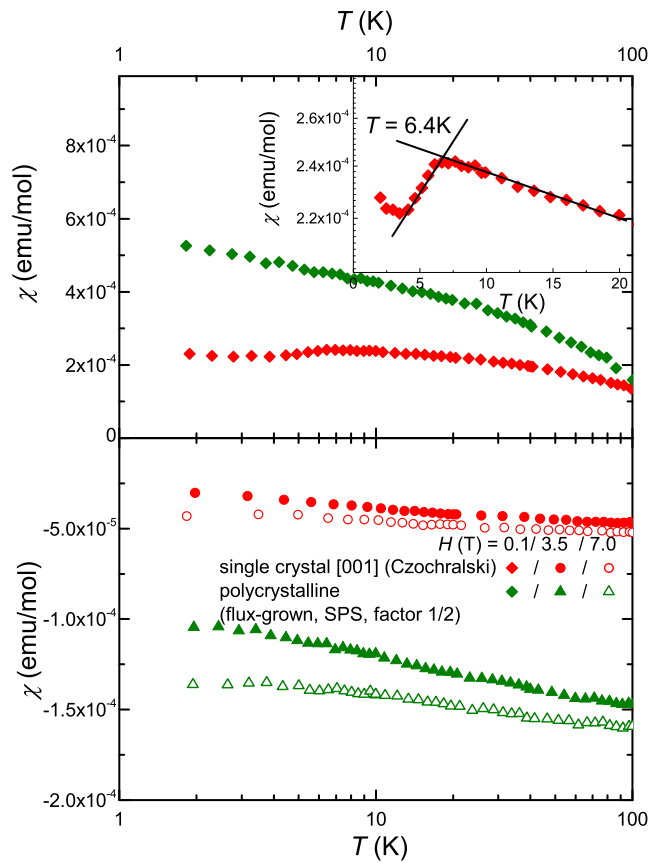}
\caption{Top: Temperature dependence of magnetic susceptibility for FeGa$_3$ single crystal (Czochralski method) oriented along [001] and 
powder made of the single crystal (flux-grown), which was additionally washed with hydrochloric acid in (Top) low external field of H 
= 0.1\,T and (Bottom) high external fields of 3.5 and 7.0\,T. The inset in the upper panel shows the magnification of temperatures below 20\,K. \label{fig:susz}}
\end{figure}

\subsection{Heat capacity and susceptibility measurements}

We proceed to investigate the likelihood of a phase transition as a source for the low temperature features in $S$ and $\kappa$. To this end, we have measured the 
heat capacity and magnetic susceptibility for FeGa$_{3}$.
The results of the heat capacity measurement is collected in Fig.~\ref{fig:Cp}(a). The zero field measurement does not
show any jumps or kinks over the entire temperature range. Moreover the data for 9\,T and low temperatures
are also coinciding with the zero-field data. This implies that there are no structural phase transitions in FeGa$_{3}$. 
The value of the electronic specific heat coefficient $\gamma$ is obtained by plotting $C_{p}$/$T$ versus $T^{2}$ (Fig.~\ref{fig:Cp}(b)) and fitting the data to $C_{p}$ = $\gamma T + \beta T^{3} + \delta T^{5}$. 
We obtained a better fit when including the $T^{5}$ term, which
 is associated to both the higher order terms of the Fourier series of a harmonic
oscillator and as well as to anharmonic terms.\cite{Passler_2013} 
This results in $\gamma$ = 0.03\,mJ\,mol$^{-1}$\,K$^{-2}$, close to zero as expected for a semiconductor and consistent with previously published values.\cite{hadano_thermoelectric_2009,umeo2012}
The Sommerfeld constant $\gamma$ in the electronic specific heat can be compared to the bare
value $\gamma_{b}$ determined from the bare reference density of states at the Fermi level, $N(0)$
($\gamma_b = \pi^{2}k_{\mathrm{B}}^{2}N(0)/3 = 2.359N(0)$),
where $N(0)$ is in states/(eV f.u.) and $\gamma_b$ is in mJ\,mol$^{-1}$\,K$^{-2}$.
From Hall effect measurements, for a carrier concentration of $n \approx$ 1.0$\times10^{20}$ electrons cm$^{-3}$
(100-300 K) and $n \approx$ 4.0$\times10^{19}$ electrons cm$^{-3}$
(< 100 K), we use the $N(0)$ from our band structure calculations and obtain
$\gamma_{b}$ = 0.06\,mJ\,mol$^{-1}$\,K$^{-2}$ and 0.01\,mJ\,mol$^{-1}$\,K$^{-2}$ respectively,
consistent with the experimental observation.

The temperature dependence of the magnetic susceptibility measured at various fields is displayed in Fig.~\ref{fig:susz}. 
For fields of 3.5\,T and  7\,T, the measured $\chi(T)$ for the [001] oriented single crystal (Czochralski method) is negative for the entire temperature range up to 300\,K
and diamagnetic. The slight upturn at low temperatures originates from the tiny amounts of Fe impurities in
the system. This fact is clearly visible in the low field 0.1\,T data where the susceptibility values at low temperatures are
positive.  For the [001] oriented single crystal (Czochralski method), in 0.1\,T field, we observe a sharp downward
turn around 6.4\,K. We were able to correlate this feature to the onset of superconductivity from tiny amounts
of Ga inclusions in thin film form ($T_{c}$ = 7.6\,K).\cite{kieselmann1985}
The gallium inclusions were removed by grinding the sample and washing it with diluted  hydrochloric acid
(H$_{2}$O:HCl = 1:1). The susceptibility measurements on the so obtained polycrystalline sample did
not show any downturn at low temperatures.
Consistent with the heat capacity measurements, we do not observe any other anomalous features in $\chi(T)$ 
that could point towards
the possibility of a magnetic phase transition. 

\subsection{Phonon-drag mechanism}

Having excluded both structural and magnetic phase transition at low temperatures, and
as well as the presence of strong electronic correlations in FeGa$_{3}$,
another possible explanation for the unusually strong signal below 20\,K in the thermopower and 
the thermal conductivity is a phonon-drag mechanism. 
The peaks in the thermal conductivity and Seebeck coefficient for all the single-crystalline samples always occur
at the same temperature, $\approx$ 20\,K. Moreover, the trend in the peak intensity between $\kappa$ and $S$ are 
similar: $\kappa_{\mathrm{polycrystalline}}$  < $\kappa_{[001]}$ < $\kappa_{[100]}$ < $\kappa_{\mathrm{unoriented}}$ 
and $S_{\mathrm{polycrystalline}}$ < $S_{[001]}$ < $S_{[100]}$ < $S_{\mathrm{unoriented}}$. 
The peak in the low temperature thermal conductivity for the flux-grown single crystal is more than 25 times
larger than for the polycrystalline sample. Similarly, the peak in the Seebeck coefficients for the flux-grown
single crystal is also larger than the powder sample (prepared from the flux-grown sample) by a factor of $\approx$ 30.  
Such features have been observed in other semiconductors including
single crystals of Ge,\cite{geballe1954} Si\cite{geballe1955} and CrSb$_{2}$,\cite{sales2012} nano-composite FeSb$_{2}$,\cite{pokharel2013} reduced TiO$_{2}$,\cite{thurber1965,tang2009} ultra-thin films of FeSi$_{2}$ and MnSi$_{1.7}$/FeSi$_{2}$,\cite{hou2011} which have 
subsequently been demonstrated to arise from phonon-drag effects. 
Phonon drag can be defined as the effect arising from a preferential 
scattering of the charge carriers by the phonons in the direction of the flow ($i.e.$, the drag on
the charge carriers exerted by the phonons streaming from hot to cold end in thermal 
conduction) and mostly evidenced at low temperatures.
With increasing temperatures, the magnitude of the phonon-drag influenced Seebeck coefficient decreases rapidly. 
This is due to the reduced relaxation time for the long wavelength phonons, which interact with the electrons/holes, as a result
of the increase in the carrier concentration. Due to this electron-phonon coupling at reduced temperatures,
the Seebeck coefficient can now be written as a sum of the conventional electron-diffusion part $S_{d}$ and
another term arising from the phonon-electron interaction $S_{p}$, the so-called phonon-drag contribution.
\begin{equation*}
S = S_{d} + S_{p}
\end{equation*}
 
 \begin{figure} [t]
\includegraphics[width=\columnwidth]{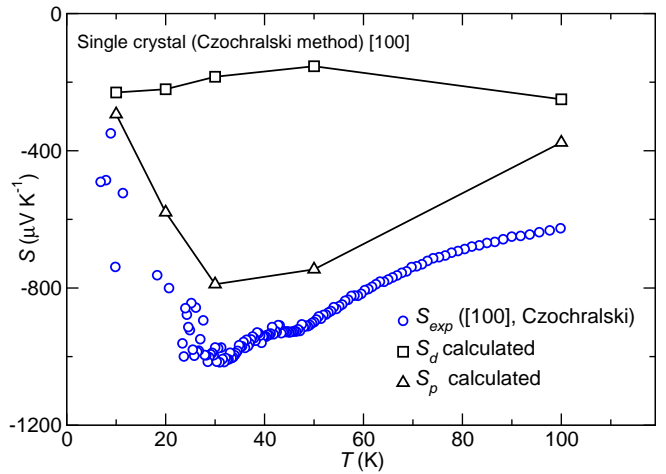}
\caption{Comparison of the measured Seebeck coefficient with the calculated values of the diffusive ($S_{d}$)
and phonon-drag ($S_{p}$) parts as a function of temperature for a single crystal of FeGa$_{3}$ (Czochralski method) along [100]. 
The $S_{p}$ term dominates the low temperature regime, indicative of a significant phonon-drag effect in 
this system. \label{fig:drag}}
\end{figure}

\begin{figure} [t]
\includegraphics[width=\columnwidth]{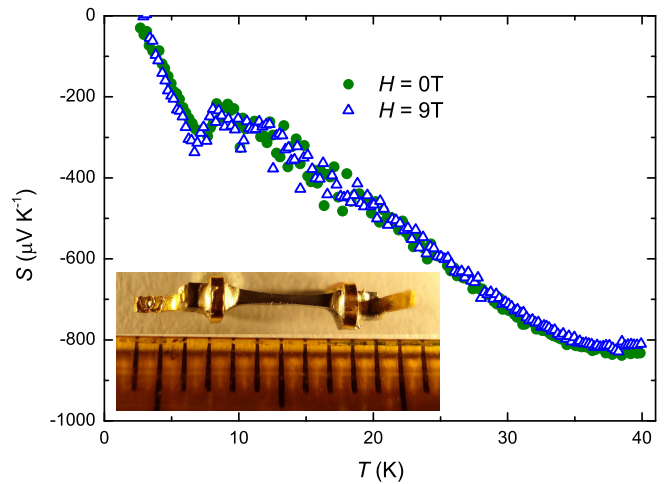}
\caption{Temperature dependence of the Seebeck coefficient for a single crystal of FeGa$_{3}$ (Czochralski method) oriented along the [100] direction, measured both at 0\,T and 9\,T magnetic fields. 
The magnetic field was applied along [100], parallel to the crystallographic orientation.
The lack of significant changes in $S$ at 
0\,T and 9\,T supports the non-electronic origin of the colossal low temperature values and thus can be
attributed to phonon-drag mechanism. The slight difference in the thermopower data compared to Fig. 5 is due to the change in the shape of the sample to improve the heat flow (see inset). \label{fig:sus-field}}
\end{figure}

Though an accurate quantitative calculation for a phonon-drag contribution is difficult to obtain, one can realize
an estimate by calculating the electron-diffusion part and subtracting it from the measured total
Seebeck coefficient.  For a semiconductor with a complex band structure and whose spin-orbit splitting is negligible ($<< k_{B}T$), the electron-diffusion part as suggested by Herring,\cite{herring1954} can be
written as
\begin{equation*}
S_{d} = \mp86.2 \left[ \ln \frac{4.70\times10^{15}}{n} + \frac{3}{2}\ln\frac{m^{*}}{m} + \frac{|\Delta\epsilon_{T}|}{k_BT} + \frac{3}{2}\ln T \right]
\end{equation*} 
where, the minus and plus signs are for the $n$ and $p$-type carriers respectively, $n$ is the charge carrier 
density denoted in cm$^{-3}$, $m$ and $m^{*}$ are the bare and inertial effective masses of the electron, respectively, |$\Delta\epsilon_{T}$| refers to the average energy of the transported electrons relative to the 
band edge, $k_B$ is the Boltzmann constant. Usually, |$\Delta\epsilon_{T}$|  is of the order of $k_BT$, and in the diffusive limit ($i.e.$, lattice scattering by phonons of long wavelength), Herring\cite{herring1954} proposes an additional approximation of |$\Delta\epsilon_{T}$|  = 5/2 + $r$ with $r$ = -1/2. Fig.~\ref{fig:drag} displays the $S_{d}$ values obtained from the above equation using $m^{*}$ = $m$ and 
$n$ from Hall resistivity measurements (Fig.~\ref{fig:hall}). Subsequently, $S_{p}$ was calculated by subtracting $S_{d}$ from the total $S$.  We note that the $S_{p}$ contribution to the total $S$ is dominant
over the electron-diffusive $S_{d}$ term in the low temperature regime and thus indicative of a significant phonon-drag effect in FeGa$_{3}$. 

Another method to eliminate the possibility of an electronic origin to the colossal low-temperature Seebeck coefficient is to measure in high magnetic fields.\cite{sales2012} If the system is dominated by the phonon-drag mechanism, only a small response in the magnetic field is expected. To confirm this scenario, we measured the Seebeck coefficient for a [100] oriented FeGa$_{3}$ single crystal (Czochralski method) at 9 T and the results are displayed in Fig.~\ref{fig:sus-field}. The lack of difference between the 0 and 9\,T data up to 40\,K is apparent and upholds the non-electronic origin of the pronounced peak in Seebeck coefficient.

\section{Summary}
We have synthesized single crystalline and polycrystalline samples of the narrow band gap semiconductor FeGa$_{3}$. 
A band gap of $\approx$ 0.5 eV is obtained from electrical resistivity measurements. Systematic investigation of the thermoelectric properties of the single- and polycrystalline samples reveal pronounced peaks around 20 K in the thermal conductivity (400-800\,W\,K$^{-1}$\,m$^{-1}$) and the Seebeck coefficient (in the order of -16000\,$\mu$V\,K$^{-1}$). Such large values have previously not been reported for FeGa$_{3}$. To identify the origin of the low temperature enhancement, we investigated various possibilities.
DFT-based band structure calculations and subsequent modeling of the Seebeck coefficient using semiclassical Boltzmann transport theory
yielded a Seebeck coefficient of the same order as the experiment above 300\,K with LDA. 
Inclusion of strong electronic correlations arising from the narrow Fe-$d$ bands treated in a mean-field manner 
(LSDA+$U$) does not improve this comparison, rendering strong correlations as an explanation for the 
low temperature enhancement in Seebeck unlikely.
Heat capacity and magnetic susceptibility measurements under different applied magnetic fields do not show any kinks or jumps around 20\,K and thus excludes a magnetic or structural phase transition as a reason for the low-temperature peak in Seebeck coefficient.
Comparing the trend in Seebeck coefficient and thermal conductivity among the various samples, and based on estimates of the phonon-drag contribution to the Seebeck coefficient and the negligible response in a high magnetic field, we conclude that the low-temperature enhancements in FeGa$_{3}$ are due to a phonon-drag effect. 

$Note\, added\, in\, revision:$ During the submission of this draft, we became aware of another recent work
on the binary FeGa$_{3}$ and the hole doped variants Fe$_{1-x}$Mn$_{x}$Ga$_{3}$ and 
FeGa$_{3-y}$Zn$_{y}$.\cite{Gamza_2014}
In contrast to our work which focusses on the anomalous thermoelectric properties
of FeGa$_{3}$ at very low temperatures (below 50\,K), 
Gam\.{z}a {\it et. al.} focus on the magnetic properties of
both the binary and the hole doped variants up to very high temperatures ($\approx$ 800\,K).\cite{Gamza_2014}
Comparison of the lattice parameters, electrical resistivity and thermodynamic 
measurements of our flux-grown
single crystals with that of Gam\.{z}a {\it et. al.}\cite{Gamza_2014} demonstrates the 
similarity between our samples. A fit to the Arrhenius law for the intrinsic region (>\,350\,K)
of the resistivity gave a band gap of 0.5 eV in our work for the single crystals grown
using the Czochralski method, while Gam\.{z}a {\it et. al.} 
obtain a value of 0.4 eV for their flux-grown samples.\cite{Gamza_2014}
Signatures of a complex magnetic ordering is presented based on neutron diffraction measurements.
In contrast, LDA+DMFT (dynamical mean field theory) calculations do not show any hint of long-range
antiferromagnetic ordering.\cite{Gamza_2014} 
However, in our work, applying LSDA+$U$ as a simple mean-field treatement of electronic
correlations, we could stabilize a simple AFM order. 
Nevertheless, this approximation did not improve the description of the thermopower.\footnote{The neutron
diffraction pattern of Gam\.{z}a {\it et. al.} (Ref.\,\onlinecite{Gamza_2014}) shows a complex
magnetic ordering in FeGa$_{3}$. Using LSDA+$U$, we have stabilized only a simple AFM order wherein 
nearest neighbor Fe sites are ordered antiferromagnetically. The discrepancy between the
measured and calculated Seebeck coefficient could then arise for two possible reasons:
the treatement of correlations in a mean-field manner is insufficient or the simple AFM 
order we consider is not a good approximation for the complex magnetic order observed in
the experiments. }

\begin{acknowledgments}
The authors acknowledge X-ray powder diffraction measurements and WDXS analysis by the competence groups Structure and Metallography at the MPI CPfS. We also thank Ralf Koban for help with physical properties measurements. D.K., H.R. and Yu.G. acknowledge funding by the DFG within SPP 1386.
\end{acknowledgments}


%

\end{document}